\documentstyle[aps,preprint]{revtex}
\def\bvec#1{\mbox{\boldmath$#1$\unboldmath}}
\def\text{\rm}
\begin{document} 
\title
{Collective T- and P- Odd Electromagnetic Moments \\ 
in Nuclei with Octupole Deformations}
\author{N. Auerbach,$^{(1)}$ V.V. Flambaum,$^{(2)}$ and
V. Spevak$^{(1)}$}
\address{$^{(1)}$School of Physics and Astronomy, Tel Aviv University,
Tel Aviv 69978, Israel}
\address{$^{(2)}$School of Physics, University of New South Wales,
Sydney, 2052, NSW, Australia}
\date{January 28, 1996}
\maketitle
\begin{abstract}
\setlength{\baselineskip}{3.5ex}
Parity and time invariance violating forces produce collective
P- and T- odd moments in nuclei with static octupole
deformation.  Collective Schiff moment, electric octupole and 
dipole and also magnetic quadrupole appear due to the mixing of
rotational levels of opposite parity and can exceed single-particle
moments by more than a factor of 100. This enhancement is due to two
factors, the collective nature of the intrinsic moments and
the small energy separation between members of parity doublets.
The above moments induce T- and P- odd effects in atoms and molecules.
Experiments with such systems may improve substantially the limits
on time reversal violation.
\\ \\
\end{abstract}
%%\pacs{24.80.+y, 32.80.Ys, 21.60.Ev, 11.30.Er}
%%
\setlength{\baselineskip}{3.ex}

Parity and time invariance nonconserving nuclear moments induced by
\mbox{P-,} T- odd nuclear forces were discussed e.g.\ in Refs.\
\cite{GFeinberg,Coveney-Sandars,Haxton-Henley,Sushkov-Flamb-Khripl,%
FKS-86,Khriplovich-book}. These moments can be enhanced in nuclei
which have close to the ground state (g.s.) levels of the same spin as
the g.s.\ but opposite parity
\cite{Haxton-Henley,Sushkov-Flamb-Khripl}. An interesting possibility
to enhance the effect is to consider mechanisms producing collective
\mbox{T-,} P- odd moments.  In Ref.\ \cite{Flambaum-spinhedg} it was   
shown that the ``spin hedgehog'' mechanism produces a collective
magnetic quadrupole.  In the present paper we want to consider a
different mechanism: mixing of opposite parity rotational levels
(parity doublets) by \mbox{T-,} P- odd interaction in the nuclei with
octupole deformation.  This deformation was demonstrated to exist in
nuclei from the Ra-Th and Ba-Sm region and produces such effects as
parity doublets, large dipole and octupole moments in the intrinsic
frame of reference and enhanced E1 and E3 transitions (see review
\cite{Ahmad-Butler}).

Let us start our consideration from the expression for the
electrostatic potential of a nucleus screened by the electrons of the
atom. If we consider only the dipole \mbox{T-,} P- odd part of
screening (Purcell-Ramsey-Schiff theorem \cite{Schiff}) one finds
\cite{Sushkov-Flamb-Khripl}: 
\begin{equation}
\varphi(\bvec{R})= \int \frac{e \rho(\bvec{r})}
{\vert \bvec{R}-\bvec{r}\vert}d^{3}\bvec{r}
+\frac{1}{Z}(\bvec{d}\bvec{\nabla})
\int \frac{\rho_{s}(r)}{\vert \bvec{R}-\bvec{r}\vert}d^{3}\bvec{r} \ .
\label{fb1} 
\end{equation}
Here $\rho(\bvec{r})$ is the nuclear charge density 
\mbox{$\int \rho(\bvec{r})d^{3}\bvec{r} =Z$}, $\rho_{s}(r)$ is
spherically symmetric part of $\rho(\bvec{r})$, and
\mbox{$\bvec{d}= \int e \bvec{r}\rho(\bvec{r})d^{3}\bvec{r}$}
is the electric dipole moment (EDM) of the nucleus.
The multipole expansion of $\varphi(\bvec{R})$ contains both 
\mbox{T-,} P- even and \mbox{T-,} P- odd terms. The dipole part in Eq.\
(\ref{fb1}) is canceled out by the second term in this  equation:
\begin{equation}
- \int e (\bvec{r}\bvec{\nabla}\frac{1}{R})\rho(\bvec{r})d^{3}\bvec{r}
+\frac{1}{Z}(\bvec{d}\bvec{\nabla})\frac{1}{R}
\int \rho_{s}(r)d^{3}\bvec{r} = 0  \ .
\label{fb2} 
\end{equation}
The next term is the quadrupole which is \mbox{T-,} P- even thus the
first non zero \mbox{T-,} P- odd term is:
\begin{eqnarray}
\varphi^{3}=&& - \frac{1}{6}
\int e \rho(\bvec{r}) r_{\alpha}r_{\beta}r_{\gamma}d^{3}\bvec{r}
\nabla_{\alpha}\nabla_{\beta}\nabla_{\gamma}\frac{1}{R}
\nonumber \\
&&
+\frac{1}{2Z}(\bvec{d}\bvec{\nabla})
\nabla_{\alpha}\nabla_{\beta}\frac{1}{R}
\int \rho_{s}(r)r_{\alpha}r_{\beta}d^{3}\bvec{r}  \ .
\label{fb3} 
\end{eqnarray}

Here \mbox{$r_{\alpha}r_{\beta}r_{\gamma}$} is a reducible tensor.
After separation of the trace there will be terms which will contain
a vector $\bvec{S}$ (Schiff) and a rank 3 
$Q_{\alpha\beta\gamma}$ (octupole) moments \cite{Sushkov-Flamb-Khripl}:
\begin{eqnarray}
&& \varphi^{3}=\varphi_{\text{Schiff}}^{3}+
\varphi_{\text{octupole}}^{3} \ ,
\nonumber \\
&& \varphi_{\text{Schiff}}^{3}=
-\bvec{S}\bvec{\nabla}\Delta\frac{1}{R}=
4\pi \bvec{S} \bvec{\nabla}\delta(R) \ ,
\nonumber \\
&&\varphi_{\text{octupole}}^{3}= 
-\frac{1}{6}Q_{\alpha\beta\gamma}
\nabla_{\alpha}\nabla_{\beta}\nabla_{\gamma}\frac{1}{R}  \ ,
\label{octup-field}
\end{eqnarray}
where
\begin{equation}
\bvec{S} = \frac{1}{10} \biggl (
\int e\rho(\bvec{r}) r^{2}\bvec{r}d^{3}\bvec{r} -
\frac{5}{3}\bvec{d}\frac{1}{Z}\int\rho_{s}(r)r^{2}
d^{3}\bvec{r} \biggr ) 
\label{fb5} 
\end{equation}
is the Schiff moment (SM) and  
\begin{eqnarray}
&&Q_{\alpha\beta\gamma}
=\!\int e\rho(\bvec{r})
[r_{\alpha}r_{\beta}r_{\gamma}-\frac{1}{5}
(\delta_{\alpha\beta}r_{\gamma} +\delta_{\beta\gamma}r_{\alpha} +     
\delta_{\alpha\gamma}r_{\beta})] d^{3}\bvec{r}
\nonumber \\
&&Q_{zzz} \equiv \frac{2}{5}Q_{3} =
\frac{2}{5}\sqrt{\frac{4\pi}{7}}
\int e\rho(\bvec{r})r^{3}Y_{30} d^{3}\bvec{r}
\label{q-zzz-y30}
\end{eqnarray}
is a tensor octupole moment. 
(Corrections to the Schiff and octupole field that arise from the fact
that we use $\rho_{s}(r)$ instead of the full $\rho(\bvec{r})$ in
Eqs.\ (\ref{fb1}-\ref{q-zzz-y30}) are of third order in the nuclear
deformation).

Here we will consider the collective SM, collective octupole and also
collective dipole as well as the collective magnetic quadrupole
resulting from the rotation of the dipole.  The mechanism for
collective SM, dipole and octupole is the following: collective
moments in the body-fixed system of the deformed nucleus are assumed
to exist without any \mbox{T-,} P- violation. However without
\mbox{T-,} P- violation the average value of these moments for a
rotational state in the laboratory system is zero.  \mbox{T-,} P- odd
mixing of rotational doublet states produces an average orientation of
the nuclear axis $\bvec{n}$ along nuclear spin $\bvec{I}$.  In the
case of a nearly degenerate rotational doublet
\begin{equation}
\Psi^{\pm}=\frac{1}{\sqrt{2}}
(\vert IMK \rangle \pm \vert IM-K \rangle) \ ,
\label{fb6} 
\end{equation}
where $K=\bvec{In}$. If the \mbox{T-,} P- interaction mixes the
members of the doublet with the coefficient $\alpha$ the total wave
function is
\mbox{$\Psi =\Psi^{+}+\alpha\Psi^{-}$} or
\begin{equation}
\Psi =\frac{1}{\sqrt{2}}
((1+\alpha)\vert IMK \rangle+ (1-\alpha)\vert IM-K \rangle) \ ,
\label{fb6-a} 
\end{equation}
one obtains
\begin{equation}
\langle \Psi \vert n_{z} \vert \Psi \rangle =
2\alpha \frac{KM}{(I+1)I} \ .
\label{fb7} 
\end{equation}
The intrinsic electric dipole and Schiff moments are directed along
$\bvec{n}$: \mbox{$\bvec{d}=d\bvec{n}$} and
\mbox{$\bvec{S}=S\bvec{n}$} and therefore have nonzero average values
in the g.s.\ \mbox{$M=K=I$}.

To elucidate the origin of collective \mbox{T-,} P- odd moments
consider a simple ``molecular'' model, shown in Fig.\ 1:
two charges $Zq$ and $q$ with masses $Z'm$ and
$m$ have coordinates \mbox{$x_{1}=-a$} and \mbox{$x_{2}=Z'a$} so 
the center of mass is at \mbox{$x=0$}.
This ``molecule'' has dipole, quadrupole, octupole and Schiff moments.
The electric dipole in this case is \mbox{$d=(Z-Z')aq$}.
Consider $Z'=Z$, in this case $d=0$, but the SM and octupole
moment are not zero.  The octupole moment $Q_{3}$ is proportional to
$Z-1$, (for $Z=1$ there 
is only quadrupole deformation). SM in the body-fixed frame is 
\begin{equation}
S=\frac{1}{10}qa^{3}Z(Z^{2}-1) \ .
\label{fb13} 
\end{equation}

For $I=\frac{1}{2}$ the SM $S \neq 0$ 
as opposed to the octupole moment which is $Q_{3} \neq 0$ 
in the intrinsic frame, but vanishes in the laboratory
because one cannot satisfy angular momentum coupling.  Thus the
octupole deformation is hidden. 
It is  possible to have a situation in which $d=0$, $Q_{3}=0$ 
in the laboratory system but the SM in the laboratory is
\mbox{$S \neq 0$}.
This result applies to any system, for example to an elementary
particle (neutron, electron). Indeed, for spin \mbox{$s=\frac{1}{2}$}
there is only one \mbox{T-,} P- odd formfactor \cite{Kobzarev,FK1980}.
However, we have shown that the two moments - EDM and SM
are not necessarily related to each other.
There is no contradiction here.
The relativistic expression for the \mbox{T-,} P- odd electromagnetic 
current for \mbox{$s=\frac{1}{2}$} in momentum representation is 
\begin{equation}
j_{\mu}=f(q^{2})\bar{\psi}\gamma_{5}\sigma_{\mu\nu}iq_{\nu}\psi \ ,
\label{fb-r1}
\end{equation}
where $q$ is the momentum transfer, $\gamma_{5}$ and 
$\sigma_{\mu\nu}$ are Dirac matrices. The formfactor can be expanded 
\begin{equation}
f(q^{2})=d+f'(0)q^{2}+...  \ ,
\label{fb-r2}
\end{equation}
where \mbox{$d=f(0)$} is the electric dipole
moment of the particle \cite{Khriplovich-book}. In
the nonrelativistic limit 
\begin{equation}
j_{0}=-f(q^{2})i \psi^{\dagger}\bvec{\sigma q}\psi \ .
\label{fb-r3}
\end{equation}
The electric potential produced by this current is 
\begin{equation}
\varphi(q^{2})\equiv j_{0}D_{0\nu}=-4\pi 
i \frac{\bvec{\sigma q}}{q^{2}} (d+f'(0)q^{2}+...) \ .
\label{fb-r4}
\end{equation}
where 
\mbox{$\displaystyle D_{\mu\nu}=(4\pi g_{\mu\nu})/q^{2}$} is the
photon propagator. In the coordinate representation 
\begin{eqnarray}
&&\bvec{q} \rightarrow i \bvec{\nabla} \ ~,~
\frac{1}{q^2} \rightarrow \frac{1}{4\pi r} \  ~,~ 
1 \rightarrow  \delta (\bvec{r}) \ ,  
\nonumber \\
&&\varphi(r)=d \bvec{\sigma \nabla}\frac{1}{r}+ 
4\pi f'(0)\bvec{\sigma \nabla}\delta(\bvec{r}) + ...  \ .
\label{fb-r5}
\end{eqnarray}
The first term in $\varphi(r)$ gives the long-range dipole field while
the second term is the contact field of the SM, i.e.\
\mbox{$\bvec{S}\sim f'(0)\bvec{\sigma}$} 
(see Eq.\ (\ref{octup-field})).  Thus, the SM emerges from the same
formfactor as the electric dipole.  (Note the difference between the
SM and the P-odd, T-even anapole moment which also produces a contact
field but corresponds to an independent formfactor {\cite{FK1980}}).
One can therefore have a priori a situation in which \mbox{T-,} P- is 
violated, the Schiff moment is not zero but the dipole moment of the
particle is zero.

The mechanism of rotational level mixing can also produce a magnetic
quadrupole.  Indeed, in the intrinsic frame of reference a deformed
nucleus can have both a magnetic dipole and magnetic quadrupole
without \mbox{T-,} P- violation. Then \mbox{T-,} P- odd interaction
mixes rotational parity doublets and can produce magnetic quadrupole
in the mixed state.
It is also worth noting that higher \mbox{T-,} P- odd moments can
appear due to rotation of lower moments. For example, rotating
electric dipole produces magnetic quadrupole.
However all these
contributions to higher moments will be proportional to
$L_{z}/(M_{_{A}}c)$ where $M_{_{A}}$ is a large mass of the nucleus
and consequently very small.

The intrinsic moments of heavy deformed nuclei are well described
using the two-fluid liquid drop model
\cite{Leander-Nazarewicz-Bertsch,Dorso-Myers-Swiatecki,%
Butler-Nazarewicz}. 
We consider here even-odd nuclei, so electric moments, except the
dipole, are determined by the moments of the even $Z$ core.
The surface of a deformed nucleus is
\begin{equation}
R=R_{0}\bigl (1+\sum_{l=1}\beta_{l}Y_{l0}\bigr ) \ .
\label{surafce} 
\end{equation}
The $\beta_{1}$ deformation is determined from the requirement that
the center of mass fixed at 
\mbox{$z=0$}, i.e.\ \mbox{$\int z d^{3}\bvec{r} =0$}:
\begin{equation} 
\beta_{1}=-3\sqrt{\frac{3}{4\pi}}
\sum_{l=2}
\frac{(l+1) \beta_{l}\beta_{l+1}}{\sqrt{(2l+1)(2l+3)}} \ .
\label{beta-one}
\end{equation}
The proton density in case of deformed nucleus is
 \cite{Leander-Nazarewicz-Bertsch}
\begin{eqnarray}
\rho=&&\frac{\rho_{0}}{2} 
-\frac{\rho_{0}}{8} \frac{e^{2}Z}{CR_{0}}
\nonumber \\
&&\times \biggl [\frac{3}{2}-\frac{1}{2}\
\Bigl ( \frac{r}{R_{0}}\Bigr )^{2}
+\sum_{l=1} \frac{3}{2l+1}\Bigl (\frac{r}{R_{0}}\Bigr )^{l}
\beta_{l}Y_{l0}\biggr ] \ ,
\label{protdis}
\end{eqnarray}
where \mbox{$\rho_{0}=3A/(4\pi R_{0}^{3})$} and
$C$ is the volume symmetry-energy coefficient of the liquid-drop
model.  The dipole moment generated by this proton distribution is in
the lowest order of deformation
\cite{Leander-Nazarewicz-Bertsch,Dorso-Myers-Swiatecki,%
Butler-Nazarewicz}
\begin{equation} 
d_{\text{intr}}=eAZ\frac{e^{2}}{C}\frac{3}{40\pi}\sum_{l=2} 
\frac{(l^{2}-1)(8l+9)}{[(2l+1)(2l+3)]^{3/2}}\beta_{l}\beta_{l+1}
\label{dipmom}
\end{equation}
The inclusion of the neutron skin effect as well as the shell
correction reduces somewhat $d_{\text{intr}}$, nevertheless Eq.\
(\ref{dipmom}) with the \mbox{$C \simeq$ 20-30 MeV} fits experimental
values quite well
\cite{Leander-Nazarewicz-Bertsch,Butler-Nazarewicz}.
This moment appears only because the Coulomb force produces a relative
shift of protons versus neutrons. The constant part of the density in  
Eq.\ (\ref{protdis}) does not contribute to $d_{\text{intr}}$.
The intrinsic Schiff moment turns out to be
\begin{equation}
S_{\text{intr}}=e AR_{0}^{3} \frac{3}{40\pi} \Bigl (
1-\frac{e^{2}Z}{R_{0}C}\frac{19}{70} \Bigr )
\sum_{l=2}
\frac{(l+1) \beta_{l}\beta_{l+1}}{\sqrt{(2l+1)(2l+3)}} \ .
\label{shiffmom}
\end{equation}
Here the constant part of $\rho$ in
Eq.\ (\ref{protdis}) gives the main contribution
(about 90\% in  nuclei with $Z\sim 90$). 
The expression for the intrinsic octupole moment is
\cite{Bohr-Mottelson} 
\begin{equation}
Q_{3~\text{intr}}=e ZR_{0}^{3}
\frac{3}{2\sqrt{7\pi}}(\beta_{3}+
\frac{2}{3}\sqrt{\frac{5}{\pi}}\beta_{2}\beta_{3} + ...) \ .
\label{oct-intr}
\end{equation}

The P- and T- odd potential has the form
\cite{Sushkov-Flamb-Khripl,Herczeg}
\begin{equation}
V^{\text{PT}}=\frac{G}{\sqrt{2}}\frac{\eta}{2m}\rho_{0}
\sum_{i}
\bvec{\sigma}_{i}
(\bvec{\nabla}_{i}f(\bvec{r}_{i})) \ ,
\label{23h}
\end{equation}
where \mbox{$G=10^{-5}/m^{2}$} is the Fermi constant and 
\mbox{$\rho_{\text{t}}(\bvec{r})=\rho_{0} f(\bvec{r})$}
is the nuclear density.

We use here the particle-core model for a reflection-asymmetric
nucleus \cite{Leander-Sheline,Brink}. The \mbox{T-,} P- odd as well as
P- odd T- even mixing was studied in this model recently \cite{SA-PLB}. 
The wave functions in the model are \cite{Leander-Sheline,SA-PLB}:
\begin{equation}
\Psi^{Ip}_{MK}=\biggl [\frac{2I+1}{16\pi^{2}}\biggr ]^{1/2}
(1+\hat{R}_{2}(\pi))D^{I}_{MK}
{\Phi}_{K}^{p} \ ,
\label{14h}
\end{equation}
where \mbox{$\hat{R}_{2}(\pi)$} denotes rotation through an angle 
$\pi$ about the intrinsic 2 axis.
The \mbox{${\Phi}^{p}\equiv {\Phi}^{\pm}$} 
are particle-core intrinsic states of good parity
$p$. Denoting the good parity core states ${\chi}^{\pi}$ and particle
states ${\phi}^{\pi}$ we write
\begin{eqnarray}
&&{\Phi}^{+}=a_{+}{\chi}^{+}{\phi}^{+} + 
b_{+}{\chi}^{-}{\phi}^{-}
\nonumber \\ 
&&{\Phi}^{-}=a_{-}{\chi}^{-}{\phi}^{+} +
b_{-}{\chi}^{+}{\phi}^{-} \ .
\label{12h}
\end{eqnarray}
The states $\chi^{\pi}$ are projections of the reflection-asymmetric
states $\chi_{_{A}}$ \cite{Leander-Sheline}
\begin{equation}
\chi^{\pi}=\frac{1}{\sqrt{2}}(1+\pi \hat{P})\chi_{_{A}}  .
\end{equation}
The matrix elements of $V^{\text{PT}}$ are \cite{SA-PLB}
\begin{eqnarray}
&&\langle \Psi^{I+}_{MK}\vert V^{\text{PT}} 
\vert \Psi^{I-}_{MK}\rangle
\nonumber \\
&&=a_{+}b_{-}
\langle \phi^{+}_{K}\vert V^{\text{PT}} \vert \phi^{-}_{K}\rangle
+a_{-}b_{+}
\langle \phi^{+}_{K}\vert V^{\text{PT}} \vert \phi^{-}_{K}\rangle \ .
\label{22ah}
\end{eqnarray}
Note that the pseudoscalar operator $V^{\text{PT}}$ cannot
connect states of an even-even axially symmetric
core $\chi^{\pi}$ \cite{SA-PLB}.
The expectation value of a \mbox{T-,} P- odd operator $\hat{O}$ in
a \mbox{T-,} P- admixed state ${\tilde{\Phi}}_{i}^{+}$ is
\begin{equation}
\langle {\tilde{\Phi}}_{i}^{+} \vert \hat{O} \vert
{\tilde{\Phi}}_{i}^{+}\rangle =
2\alpha_{ii} \langle {\Phi}_{i}^{+}\vert \hat{O} \vert
{\Phi}_{i}^{-} \rangle  +
2\sum_{j\neq i}\alpha_{ij}
\langle {\Phi}_{i}^{+}\vert \hat{O} \vert
{\Phi}_{j}^{-} \rangle  \ .
\end{equation}
The matrix elements between core states are
\begin{equation}
\langle \chi^{+} \vert \hat{O} \vert \chi^{-} \rangle=
\langle \chi_{_{A}} \vert \hat{O} \vert \chi_{_{A}} \rangle \ .
\end{equation}
Writing the one-body operator $\hat{O}$ as sum of core and particle
parts \mbox{$\hat{O}=\hat{O}_{\text{core}}+\hat{O}_{\text{p}}$} 
one obtains 
\begin{eqnarray}
\langle {\Phi}_{i}^{+}\vert \hat{O} \vert
{\Phi}_{j}^{-} \rangle=
&&
\langle \chi_{_{A}} \vert \hat{O}_{\text{core}} \vert 
\chi_{_{A}} \rangle (a_{{+}~{i}}a_{{-}~{j}}+b_{{+}~{i}}b_{{-}~{j}})
\nonumber \\
&&
+\langle {\phi}^{+}_{i} \vert \hat{O}_{\text{p}} \vert 
{\phi}^{-}_{j} \rangle 
(a_{{+}~{i}}b_{{-}~{j}}+a_{{-}~{i}}b_{{+}~{j}})
\end{eqnarray}
The contribution of the single neutron is small for
the SM operator and is absent for the octupole moment.
In the case of closely spaced doublets 
\mbox{$a_{+~i}\approx a_{-~i}$}, \mbox{$b_{+~i}\approx b_{-~i}$}
and
\mbox{$(a_{{+}~{i}}a_{{-}~{j}}+b_{{+}~{i}}b_{{-}~{j}})\approx
\delta_{ij}$}.
The expressions for the expectation values of a \mbox{T-,} P- odd
operator of rank $l$ in the body-fixed and laboratory (for the g.s.\
\mbox{$I=M=K$} \cite{Bohr-Mottelson}) systems becomes
\begin{eqnarray}
&&\langle {\tilde{\Phi}}_{i}^{+} \vert \hat{O} \vert 
{\tilde{\Phi}}_{i}^{+}\rangle \approx
2\alpha_{ii}
\langle \chi_{_{A}} \vert \hat{O}_{\text{core}} \vert 
\chi_{_{A}} \rangle \ ,
\nonumber \\
&&\langle \tilde{\Psi}^{I+}_{MK}\vert\hat{O}\vert
\tilde{\Psi}^{I+}_{MK} \rangle =
\langle II l0 \vert II \rangle^{2}
\langle {\tilde{\Phi}}_{i}^{+} \vert \hat{O} \vert 
{\tilde{\Phi}}_{i}^{+}\rangle \ .
\label{mom-expvalues}
\end{eqnarray}

Currently, the best limits on Schiff moments and the coupling
constants of \mbox{T-,} P- violating nucleon-nucleon interactions are
obtained from the measurements of the electric dipole moments in
$^{199}$Hg, $^{129}${Xe} \cite{d-Hg-95} atoms and TlF molecule
\cite{d-TlF-89,Barr-review}.  Nuclei of these atoms do not have
octupole deformation. However, similar experiments can be done with
heavy atoms (Ra, Rn) which are electronic structure analogs of these
atoms but their nuclei have octupole deformation.

Our calculations were performed for relatively long lived even-odd
isotopes $^{223,225}$Ra and $^{223}$Rn.  Variants of the model used
here are shown to describe quite well the g.s.\ parity doublets 
in the Ra-Th region
\cite{Leander-Sheline,Cwiok-Nazarewicz,Ahmad-Butler}.
We used here the same version as in Ref. \cite{SA-PLB},
the deformation and core parity splitting parameters were taken from   
Ref.\ \cite{Leander-Sheline,Cwiok-Nazarewicz}. The calculations 
of the mixing coefficients were performed using Nilsson potential.
The $^{223,225}$Ra and $^{223}$Rn have the g.s.\ 
$\frac{3}{2}^{+}$, $\frac{1}{2}^{+}$ and $\frac{7}{2}$.
The octupole deformation for all these isotopes is
$\beta_{3}\approx 0.1$. 
Our results are shown in table \ref{moments-table}.
The mixing coefficients in our calculations  are
in the range \mbox{(0.6--7.)$\times 10^{-7}\eta$}.
The Schiff moments of the reflection asymmetric nuclei
in the intrinsic system are in the range 
\mbox{(22.--29.)$e\text{fm}^{3}$}.
Correspondingly the \mbox{T-,} P- odd Schiff moments 
(Eq.\ (\ref{mom-expvalues}))
are in the range 
\mbox{(2.--20.)$\times 10^{-6}\eta e\text{fm}^{3}$}.
This is about two orders of magnitude larger than the largest
single-particle estimate of
\mbox{4.$\times 10^{-8}\eta e\text{fm}^{3}$} given in 
Ref. \cite{Sushkov-Flamb-Khripl} for $^{237}$Np.

Since the electronic structure of Ra is similar to that of Hg and
Rn is similar to Xe, we can use results of atomic
calculations for Hg, Xe \cite{Dzuba-Flambaum,FKS-86} and write
\begin{eqnarray}
&&d_{\text{at}}({\text{Ra}})=d_{\text{at}}({\text{Hg}})
\frac{(SZ^{2}R_{1/2})_{\text{Ra}}}{(SZ^{2}R_{1/2})_{\text{Hg}}} \ ,
\nonumber \\
&&d_{\text{at}}({\text{Rn}})=d_{\text{at}}({\text{Xe}})
\frac{(SZ^{2}R_{1/2})_{\text{Rn}}}{(SZ^{2}R_{1/2})_{\text{Xe}}} \ ,
\label{fb-A1} 
\end{eqnarray}
where we have taken into account the Z- dependence of the Schiff
moment contribution to the atomic EDM $d_{\text{at}}$ due to increase
of electronic wave functions near the nucleus
\cite{Sushkov-Flamb-Khripl} \mbox{$d_{\text{at}} \sim SZ^{2}R_{1/2}$}.
The relativistic factor is given by 
\begin{equation}
R_{1/2}=\frac{4\gamma_{1/2}}{[\Gamma(2\gamma_{1/2}+1)]^{2}}
\biggl (\frac{2ZR_{0}}{a_{_{B}}} \biggr )^{2\gamma_{1/2}-2} \ ,
\label{fb-at1}
\end{equation}
where \mbox{$\gamma_{1/2}=[1-(Z\alpha)^{2}]^{1/2}$}, $a_{_{B}}$ is the
Bohr radius and $R_{0}$ is the nuclear radius.

We made an estimate also for $^{229}$Pa which has the smallest energy
splitting between members of g.s.\ parity doublet among the known
octupole deformed nuclei \cite{Ahmad-Butler}.  We assumed the same
atomic physics parameters as for Ra, Rn.  As seen in table
\ref{moments-table} the EDM of Ra and Rn are \mbox{$10^{2}$-$10^{3}$}
times larger than EDM of Hg and Xe.  
We should stress that the phenomenon described in our work occurs
probably in many other nuclei that possess octupole deformation (for
example in the Ba-Sm region) and which might be more suitable choice
for experimental studies. Experiments with atoms or molecules
containing these nuclei may improve substantially the limits on time
reversal violation.

This work was supported by the US-Israel Binational Science Foundation
and by a grant for Basic Research of the Israel Academy of Science.
One of us (V.V.F.) would like to thank Tel Aviv University for
hospitality and M.Yu.~Kuchiev for useful discussions.

%%%%%%%%%%%%%%%%%%%%
\begin{table}
\caption{Admixture coefficients $\alpha$ (absolute values),
experimental energy splitting between the g.s.\ doublet levels 
\mbox{$\Delta E=E^{-}-E^{+}$},
{\em intrinsic} Schiff moments and Schiff moments 
as well as induced atomic dipole moments.
The values for $^{199}$Hg and $^{129}$Xe from 
 Ref.\ \protect\cite{Dzuba-Flambaum,FKS-86}
are given for comparison.}
\begin{tabular}{lcccccc}
 &$^{223}$Ra&$^{225}$Ra&$^{223}$Rn&$^{229}$Pa&$^{199}$Hg&$^{129}$Xe\\
\hline
$\alpha
\times 10^{7}[\eta]$& 2.  & 6.  & 2.  & 60.  &    \\
$\Delta E$ [keV] & 50.2&  31.6& 130.\tablenotemark[1]&  0.22 &  & \\
$S_{\text{intr}}
[e~\text{fm}^{3}]$&  22.   & 29.    &   22.   & 28.   &  &  \\
$\mbox{$S \times 10^{8}$} \atop
\mbox{$[\eta~e~\text{fm}^{3}]$}$& 500.& 1100.& 700.& 
3.$\times 10^{5}$&     -1.4&      1.75 \\    
$\mbox{$d(\text{at})
\times 10^{25}$} \atop
\mbox{$[\eta~ e~\text{cm}]$}$& 3500.& 7900.& 1500.& 
3.$\times 10^{5}$&       5.6&     0.47  \\ 
\end{tabular}
\label{moments-table}
\tablenotemark[1]{Calculated}
\end{table}   
%%%%%%%%%%%%%%%%%%%%
%%%  The figure is produced using {picture}
%%%%%%%%%%%%%%%%%
\vspace*{10mm}
\centerline{FIGURES}
\small
%%%
\setlength{\unitlength=1mm}
\thinlines
\begin{center}
\begin{picture}(50,9)(-25,-3)
\put(0,0){\line(1,0){30}}
\put(5,2){$L=a(Z'+1)$}
\put(29.2,0){\circle*{1.6}}
\put(33.,0){$m$, $q$}
%%%
\put(-1.5,0){\circle*{3.0}}
\put(-18.,0){$Z'm$, $Zq$}
\end{picture}
\end{center}
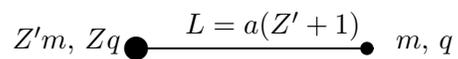
{~~FIGURE 1.}~~A ``molecular'' model of octupole deformation: two charges
$Zq$ and $q$ with masses $Z'm$ and $m$ 
placed at $-a$ and $Z'a$ with respect to the center of mass.
%%%%%%%%%%%%%%%%%%

\begin{thebibliography}{999}
\bibitem{GFeinberg}G. Feinberg,
Trans.\ N.-Y.\ Ac.\ Sc., Ser II, {\bf 38}, 26 (1977).
\bibitem{Coveney-Sandars}P.V. Convey, P.G.H.~Sandars, 
J.\ Phys.\ B {\bf 16}, 3727 (1983). 
\bibitem{Haxton-Henley}W.C. Haxton and E.M.~Henley,
Phys.\ Rev.\ Lett. {\bf 51}, 1937 (1983).
\bibitem{Sushkov-Flamb-Khripl}O.P.~Sushkov, V.V.~Flambaum, and
I.B.~Khriplovich, Sov.\ Phys.\ JETP {\bf 60}, 873 (1984).
\bibitem{FKS-86}V.V.~Flambaum, I.B.~Khriplovich and O.P.~Sushkov,
Nucl.\ Phys. {\bf A449}, 750 (1986).
\bibitem{Khriplovich-book}I.B.~Khriplovich, Parity Nonconservation in
Atomic Phenomena (Gordon and Breach, New York, 1991). 
\bibitem{Flambaum-spinhedg}V.V.~Flambaum,
Phys.\ Lett.\ B {\bf 320}, 211 (1994).
\bibitem{Ahmad-Butler}I. Ahmad and P.A.~Butler,
Annu.\ Rev.\ Nucl.\ Part.\ Sci. {\bf 43}, 71 (1993).
\bibitem{Schiff}L.I. Schiff, 
Phys.\ Rev. {\bf 132}, 2194 (1963).
\bibitem{Kobzarev}I.Yu. Kobzarev, L.B. Okun, M.V. Terent'ev,
JETP Lett. {\bf 2}, 289 (1965).
\bibitem{FK1980}V.V.~Flambaum and I.B.~Khriplovich,
Sov.\ Phys.\ JETP {\bf 52}, 835 (1980).
\bibitem{Leander-Nazarewicz-Bertsch}G.A. Leander, W.~Nazarewicz,
G.F.~Bertsch, J.~Dudek,
Nucl.\ Phys. {\bf A453}, 58 (1986), and references therein.
\bibitem{Dorso-Myers-Swiatecki}C.O.~Dorso, W.D.~Myers and W.J.~Swiatecki,
Nucl.\ Phys. {\bf A451}, 189 (1986).
\bibitem{Butler-Nazarewicz}P.A. Butler and W.~Nazarewicz,
Nucl.\ Phys. {\bf A533}, 249 (1991).
\bibitem{Bohr-Mottelson}A. Bohr and B.~Mottelson, {Nuclear Structure}, 
Vol.\ 2 (Benjamin, New York, 1975).
\bibitem{Herczeg}P. Herczeg, Hyperfine Interact. {\bf 43} (1988) 127.
\bibitem{Leander-Sheline}G.A.~Leander and R.K.~Sheline,
Nucl.\ Phys. {\bf A413}, 375 (1984) ;
G.A. Leander and Y.S.~Chen, 
Phys.\ Rev.\ C {\bf 37}, 2744 (1988).
\bibitem{Brink}D.M. Brink {et al.},
J.\ Phys.\ G {\bf 13}, 629 (1987).
\bibitem{SA-PLB}V.~Spevak and N.~Auerbach,
Phys.\ Lett.\ B {\bf 359}, 254 (1995);
N.~Auerbach, J.D.~Bowman, and V.~Spevak,
Phys.\ Rev.\ Lett. {\bf 74}, 2638  (1995).
\bibitem{d-Hg-95}J.P.~Jackobs, W.M.~Klipstein, S.K.~Lamoreaux,
B.R.~Heckel, E.N.~Fortson,
Phys.\ Rev.\ A {\bf 52}, 3521 (1995); 
T.G.~Vold {\em et al.},
Phys.\ Rev.\ Lett. {\bf 52}, 2229 (1984).
\bibitem{d-TlF-89}D. Cho, K.~Sangster, E.A.~Hinds,
Phys.\ Rev.\ Lett. {\bf 63}, 2259 (1989).
\bibitem{Barr-review}S.M.~Barr,
Int.\ J.\ Mod.\ Phys. A {\bf 8}, 209 (1993).
\bibitem{Cwiok-Nazarewicz}S.~\'{C}wiok, W.~Nazarewicz,
Nucl.\ Phys. {\bf A529}, 95 (1991).
\bibitem{Dzuba-Flambaum}V.A.~Dzuba, V.V.~Flambaum, P.G.~Silvestrov,
Phys.\ Lett.\ B {\bf 154}, 93 (1985).
\end{thebibliography}
\end{document}